\documentclass[eqsecnum]{revtex4}
\usepackage{epsfig,amsmath}

\def\l{\langle}
\def\r{\rangle}
\def\be{\begin{eqnarray}}
\def\ee{\end{eqnarray}}
\def\a{\alpha}
\def\bo{\begin{figure}}
\def\eo{\end{figure}}

\begin{document}

\title{Multiparticle Entanglement with Quantum Logic Networks:  
Application to Cold Trapped Ions}

\author{Marek \v Sa\v sura$^{1}$ and Vladim{\'\i}r Bu\v zek$^{1,2}$}
\address{
$^{1}$ Institute of Physics, Slovak Academy 
of Sciences, D\'{u}bravsk\'{a} cesta 9, Bratislava 842 28, Slovakia
\\
$^{2}$
Faculty of Informatics, Masaryk University, Botanick\'a 68a, Brno 602 00,
Czech Republic    
}
\date{February 5, 2001}

\begin{abstract}
We show how to construct a multi-qubit control gate on a quantum register
of an arbitrary  size $N$.
This gate performs a single-qubit operation on a specific qubit conditioned
by the state of other $N-1$ qubits.
We provide an algorithm how to build up an array of networks
consisting of single-qubit rotations and multi-qubit control-NOT
gates for the
synthesis of an arbitrary entangled quantum state of $N$ qubits.  We
illustrate the algorithm on a system of cold trapped ions. This example
illuminates the efficiency of the direct implementation of the multi-qubit
CNOT gate compared to its decomposition into a network of 
two-qubit CNOT gates. \newline
\end{abstract}

\maketitle

\section{Introduction}
Entanglement is probably the most intriguing aspect of quantum
theory \cite{Schroedinger}. It attracts due attention not only 
for its epistemological importance \cite{Peres} but also 
as an essential resource for
quantum information processing. In particular,
quantum computation \cite{Nielsen2000, Gruska1999},
quantum teleportation \cite{Bennett93}, quantum dense coding
\cite{Bennett92},  certain types of quantum key distributions
\cite{Ekert91} and quantum secret sharing protocols \cite{Hillery99},
are rooted in the existence of quantum entanglement. 

Recently, lot of progress has been achieved in investigation of various
properties and possible application of quantum entanglement. Nevertheless,
many questions are still opened. In particular, it is the problem of
multi-particle entanglement \cite{Thapliyal99}. Specifically, in contrast
to classical correlations, quantum entanglement cannot be freely shared among
many objects \cite{Coffman,Wootters}. It has been shown recently
\cite{Dur, Koashi} that in a finite system of $N$ qubits with $N(N-1)/2$
entangled pairs the maximal possible concurrence (a specific measure of
entanglement \cite{Wootters, Hill}) is equal to $2/N$. This value
of the bipartite concurrence is achieved when the $N$ qubits are prepared
in a totally symmetric state $|\Xi\r$, such that all except one qubit
are in the state $|1\r$, i.e.
\be
\label{1.1}
|\Xi\r=\frac{1}{\sqrt{N}}\sum_{j=1}^N|0\r_j|1\r^{N-1}
=\frac{1}{\sqrt{N}}\bigg(
|011\dots 1\r+|101\dots 1\r+|110\dots 1\r+\dots+|111\dots 0\r
\bigg)\, .
\ee  
In order to study the multiparticle 
quantum entanglement in more detail, we have to find
ways how to prepare (synthesize)  states of the form given by
Eq.~(\ref{1.1}) in various physical systems.

In this paper we will study in detail how $N$ qubits
can be prepared in entangled states of the form (\ref{1.1}).
We assume that
the qubits are encoded in internal ionic states as originally proposed
in the model of quantum processor by Cirac and Zoller \cite{c&z}.
Our paper is organized as follows: Section~\ref{sec2} is devoted
to the description of quantum logic gates and networks. Here we present
multi-qubit controlled gates.  
We show how these gates can be  expressed in terms of single-qubit and 
two-qubit gates, but we argue that for practical purposes it is more
appropriate to utilize directly multiple-qubit gates rather than decompose
them into elementary single and two-qubit gates.

In Section~\ref{sec3} we present a logical network with the help of
which symmetric states of the form (\ref{1.1}) can be synthesized
Section~\ref{sec4} is devoted to a general problem of synthesis of a pure
state of an arbitrary $N$ qubit state. We present a simple network using
which an arbitrary $N$ qubit state can be created. In
Section~\ref{sec5} we apply this algorithm to a specific problem of
$N$ cold trapped ions. Following the original idea of Cirac and Zoller
we show how the states of interest can be created. In the last 
Section ~\ref{sec6} we discuss the experimental realization of the proposed
scheme on cold trapped ions and we also briefly address the efficiency of
using multi-qubit control-NOT gates rather than a network of two-qubit
control-NOT gates.

\section{Quantum logic gates and networks}
\label{sec2}
Let us start  with a brief description of those objects we will 
use later in the paper.
We will follow
the notation used in Ref.~\cite{ekert, Nielsen2000}.  
The {\it qubit} (quantum bit) is a quantum two-level system in which
logical Boolean states 0 and 1 are represented by a pair of normalized and
mutually orthogonal quantum states labelled as $|0\r$ and $|1\r$. These two
states form a computational basis and any other pure state of the qubit can
be written as a coherent superposition $|\psi\r=\alpha|0\r+\beta|1\r$ 
with complex amplitudes 
$\alpha$ and $\beta$, such that $|\alpha|^2+|\beta|^2=1$. We may
represent a state of a qubit as a point on the Bloch sphere 
with the parameterization 
$\alpha=\cos\vartheta/2$ and $\beta=e^{i\varphi}\sin\vartheta/2$. In quantum
or atomic optics the qubit is often represented by a two-level atom
(ion) with two selected internal levels denoted  as $|g\r$ and $|e\r$. 
The {\it quantum register} of
size $N$ is a collection of $N$ qubits. The {\it quantum logic gate} is a
quantum device which performs a  unitary operation on selected 
(target) qubits conditioned by states of control qubits during a given
interval 
of time. A gate acting on a single qubit is termed as a  single-qubit
gate, gates acting on more qubits are referred to as  multi-qubit gates.
The {\it quantum logic network} is a quantum device consisting of several 
quantum logic gates synchronized in time. 

\subsection{Single-qubit rotation}

A   single-qubit gate corresponds to a unitary 
operator $W$ represented in the computational basis
$\{|0\r,|1\r\}$ by the matrix
\be
\label{2.2}
W=\left(
\begin{array}{rr}
W_{00} & W_{01}\\\
W_{10} & W_{11}
\end{array}
\right)\,.
\ee
A special case of a  single-qubit gate is a single-qubit rotation
$O$  [see FIG. \ref{obr1} (a)].
Its parameterization depends on the choice of 
coordinates on the Bloch sphere. We will define it in the matrix form in the
basis $\{|0\r,|1\r\}$ as follows
\be
\label{2.3}
O(\theta,\phi)=
\left(
\begin{array}{cc}
R_{00} & R_{01}\\
R_{10} & R_{11}
\end{array}
\right)=
\left(
\begin{array}{cc}
\cos(\theta/2) & e^{i\phi}\sin(\theta/2)\\
-e^{-i\phi}\sin(\theta/2) & \cos(\theta/2)
\end{array}
\right)\,,
\ee
where $\theta$ refers to the rotation and $\phi$ to the relative phase shift
of the states $|0\r$ and $|1\r$ in the corresponding Hilbert space.

\bo[htb]
\centerline{\epsfig{width=6cm, file=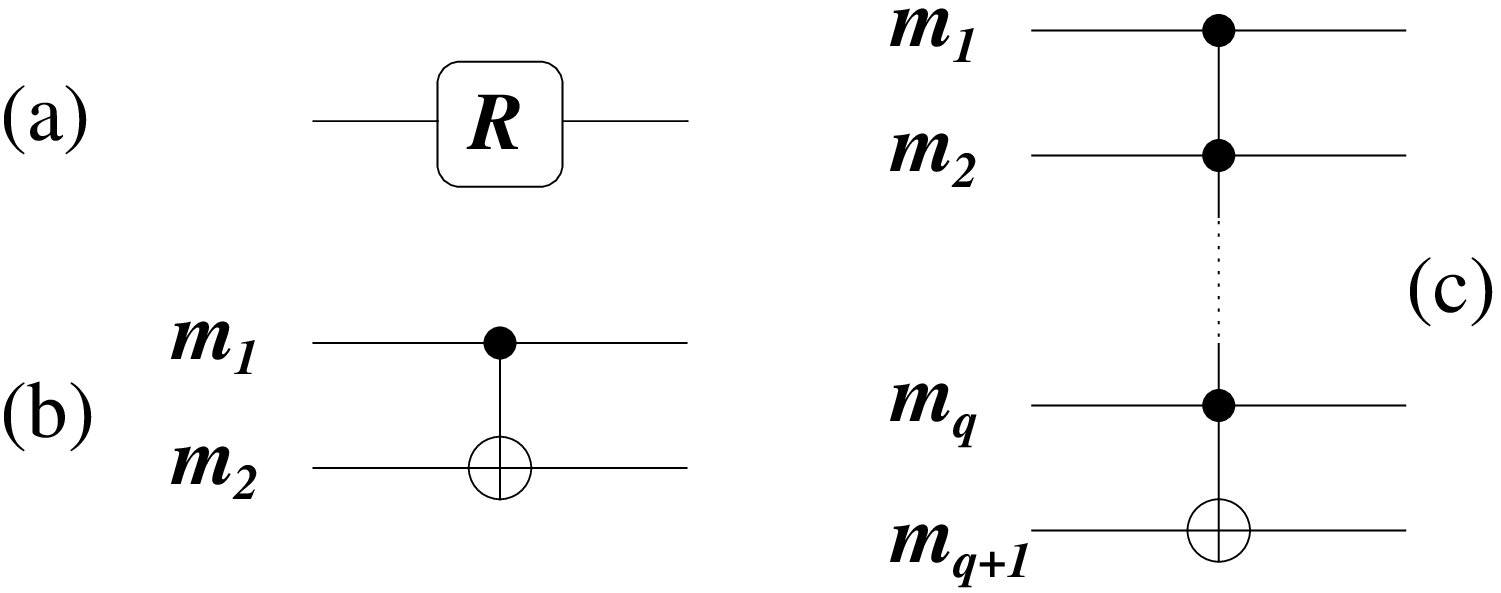}}
\caption{{\footnotesize
A schematic representation of (a) a single-qubit rotation
defined by the relation (\ref{2.3}), (b) a  two-qubit CNOT gate 
defined by the transformation (\ref{2.4}) and (c) a  
multi-qubit  $(\mbox{control})^q$-NOT gate defined by the transformation
(\ref{2.5}).}}
\label{obr1}
\eo

\subsection{Two-qubit and multi-qubit control-NOT gates}

A two-qubit control-NOT (CNOT) gate acts on two quantum bits denoted as 
the control and the target qubit, respectively [see FIG.~\ref{obr1} (b)].
If the control qubit ($m_1$) is in the state $|1\r$, the
state of the target qubit ($m_2$) is flipped. 
Otherwise, the gate acts trivially, i.e.
as a unity operator $\openone$.
We may characterize this gate with the help of the truth table
\be
\label{2.4}
\begin{array}{lll}
|0\r_{m_1}|0\r_{m_2} & \longrightarrow & |0\r_{m_1}|0\r_{m_2}\,,\\
|0\r_{m_1}|1\r_{m_2} & \longrightarrow & |0\r_{m_1}|1\r_{m_2}\,,\\
|1\r_{m_1}|0\r_{m_2} & \longrightarrow & |1\r_{m_1}|1\r_{m_2}\,,\\
|1\r_{m_1}|1\r_{m_2} & \longrightarrow & |1\r_{m_1}|0\r_{m_2}\,.
\end{array}
\ee

A multi-qubit control-NOT (CNOT) gate is defined analogically 
[see FIG. \ref{obr1} (c)]. The only difference is the
number of control qubits. In other words, a multi-qubit $(\mbox{control})^q$-NOT
gate acts on $q+1$ qubits with $q$ control qubits ($m_1,\dots,m_q$)
and the $m_{q+1}$ qubit is target. If all
control qubits are  in the state $|1\r$, then the state of the
target qubit is flipped. Otherwise, the gate action is trivial. The truth
table of the multi-qubit $(\mbox{control})^q$-NOT gate acting on
$m_1,\dots,m_{q+1}$ qubits reads as follows 
\be
\label{2.5}
\begin{array}{llll}
|\Psi_{no}\r|0\r_{m_{q+1}} & \longrightarrow & 
\quad |\Psi_{no}\r|0\r_{m_{q+1}}\,, &
\quad |\Psi_{no}\r\neq\prod\limits_{j=1}^q\otimes|1\r_{m_j}\,,\\
|\Psi_{no}\r|1\r_{m_{q+1}} & \longrightarrow & 
\quad |\Psi_{no}\r|1\r_{m_{q+1}}\,, & \\
|\Psi_{yes}\r|0\r_{m_{q+1}} & \longrightarrow & 
\quad |\Psi_{yes}\r|1\r_{m_{q+1}}\,, &
\quad |\Psi_{yes}\r=\prod\limits_{j=1}^q\otimes|1\r_{m_j}\,,\\
|\Psi_{yes}\r|1\r_{m_{q+1}} & \longrightarrow & 
\quad |\Psi_{yes}\r|0\r_{m_{q+1}}\,. &
\end{array}
\ee

\subsection{Multi-qubit control-$R$ gates}

A multi-qubit $(\mbox{control})^q$-$R$ gate acts  on $q+1$ qubits. 
The $m_1,\dots,m_q$ qubits represent the control part of the gate
while  the $m_{q+1}$ qubit represents the target 
[FIG. \ref{obr2}]. 
This gate   performs a single-qubit
rotation (\ref{2.3}) on the target qubit if all control qubits are in the state
$|1\r$. Otherwise, it acts trivially. Speaking precisely, if all control
qubits ($m_1,\dots,m_q$) are in the state $|1\r$, then the operation 
$R=R_1^{\dag}\,\sigma\,R_2^{\dag}\,\sigma\,R_2\,R_1$ is applied 
(from right to left) on the
$m_{q+1}$ (target) qubit. In the basis of the target qubit 
$\{|0\r_{m_{q+1}},|1\r_{m_{q+1}}\}$ we can introduce the matrices
\be
\label{2.6}
&R=
\left(
\begin{array}{cc}
\cos\theta & e^{i2\phi}\sin\theta\\
-e^{-i2\phi}\sin\theta & \cos\theta
\end{array}
\right)\,,\qquad
&\sigma=
\left(
\begin{array}{cc}
0 & 1 \\
1 & 0
\end{array}
\right)\,,
\nonumber
\\
&R_1= \left(
\begin{array}{cc}
0 & e^{i\phi}\\
-e^{-i\phi} & 0
\end{array}
\right)\,,\qquad
&R_1^{\dag}=
\left(
\begin{array}{cc}
0 & -e^{i\phi} \\
e^{-i\phi} & 0
\end{array}
\right)\,,
\nonumber\\
\label{2.8}
&R_2=
\left(
\begin{array}{rc}
\cos(\theta/2) & \sin(\theta/2) \\
-\sin(\theta/2) & \cos(\theta/2)
\end{array}
\right)\,,\qquad
&R_2^{\dag}=
\left(
\begin{array}{cr}
\cos(\theta/2) & -\sin(\theta/2) \\
\sin(\theta/2) & \cos(\theta/2)
\end{array}
\right)\,,
\ee
where $R_1=O(\pi,\phi)$, $R_1^{\dag}=O^{\dag}(\pi,\phi)$, 
$R_2=O(\theta,0)$ and $R_2^{\dag}=O(\theta,0)$. The operation
$O(\theta,\phi)$ is defined by the relation (\ref{2.3}). The matrix $\sigma$
denotes the NOT operation. If not all control qubits are in the state $|1\r$,
then the gate performs on the target qubit the operation 
$\openone=R_1^{\dag}\,\openone\,R_2^{\dag}\,\openone\,R_2\,R_1$, 
where $\openone$ is the unity operator.
We may write the truth table of the multi-qubit $(\mbox{control})^q$-$R$
gate as follows 
\be
\label{2.9}
\begin{array}{lll}
|\Psi_{no}\r|0\r_{m_{q+1}} & \longrightarrow & 
\quad |\Psi_{no}\r|0\r_{m_{q+1}}\,,\\
|\Psi_{no}\r|1\r_{m_{q+1}} & \longrightarrow & 
\quad |\Psi_{no}\r|1\r_{m_{q+1}}\,,\\
|\Psi_{yes}\r|0\r_{m_{q+1}} & \longrightarrow & 
\quad |\Psi_{yes}\r
\big(\cos\theta\,|0\r_{m_{q+1}}-e^{-i2\phi}\sin\theta\,|1\r_{m_{q+1}}\big)\,,
\\
|\Psi_{yes}\r|1\r_{m_{q+1}} & \longrightarrow & 
\quad |\Psi_{yes}\r
\big(e^{i2\phi}\sin\theta\,|0\r_{m_{q+1}}+\cos\theta\,|1\r_{m_{q+1}}\big)\,,
\end{array}
\ee
where $|\Psi_{no}\r$ and $|\Psi_{yes}\r$ are defined in (\ref{2.5}).

\bo[htb] 
\centerline{\epsfig{width=8cm, file=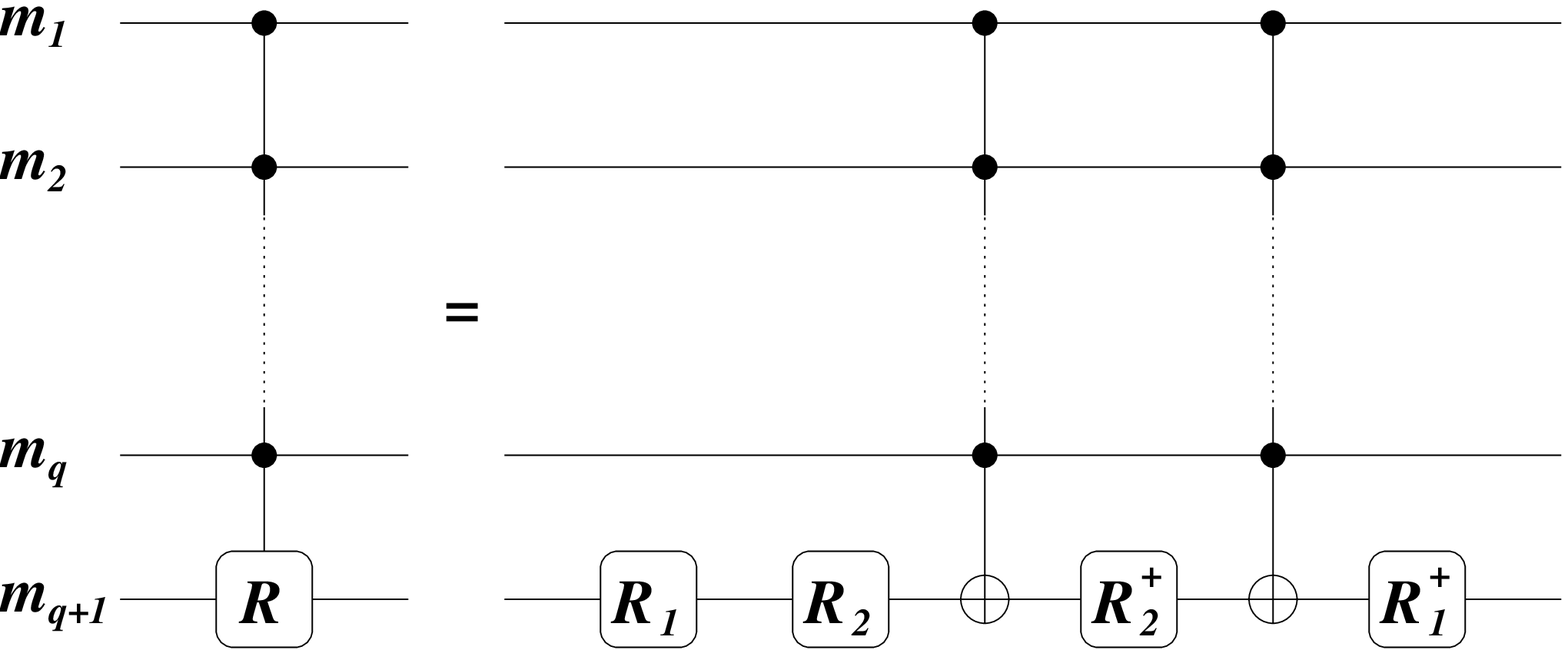}}
\caption{{\footnotesize
A scheme of a  multi-qubit $(\mbox{control})^q$-NOT gate acting
on $q+1$ qubits with $q$ control qubits ($m_1,\dots,m_q$).
The
$m_{q+1}$th qubit is the target. 
The operators $R, R_1, R_2, R_2^{\dag}$ and $R_1^{\dag}$ 
are defined by Eq.~(\ref{2.6}) and the gate
is determined by the transformation (\ref{2.9}). 
The gate corresponding to Eq.~(\ref{2.10}) is
represented by the same network except 
the single-qubit rotations $R_1$ and $R_1^{\dag}$.}}
\label{obr2}
\eo

If the preparation of a particular class of quantum states does not
require the introduction of a relative phase shift $\phi$ between the basis
states $|0\r$ and $|1\r$, then a reduced quantum logic network is 
sufficient. In particular, 
the operation $R=\sigma\,R_2^{\dag}\,\sigma\,R_2$ on
the target qubit ($m_{q+1}$)
conditioned by the state of control qubits ($m_1,\dots,m_q$) 
can be realized according to the following truth table
\be
\label{2.10}
\begin{array}{lll}
|\Psi_{no}\r|0\r_{m_{q+1}} & \longrightarrow & 
\quad |\Psi_{no}\r|0\r_{m_{q+1}}\,,\\
|\Psi_{no}\r|1\r_{m_{q+1}} & \longrightarrow & 
\quad |\Psi_{no}\r|1\r_{m_{q+1}}\,,\\
|\Psi_{yes}\r|0\r_{m_{q+1}} & \longrightarrow & 
\quad |\Psi_{yes}\r
\big(\cos\theta\,|0\r_{m_{q+1}}-\sin\theta\,|1\r_{m_{q+1}}\big)\,,\\
|\Psi_{yes}\r|1\r_{m_{q+1}} & \longrightarrow & 
\quad |\Psi_{yes}\r
\big(\sin\theta\,|0\r_{m_{q+1}}+\cos\theta\,|1\r_{m_{q+1}}\big)\,.
\end{array}
\ee

The results given above for the multi-qubit control-$R$ gates are 
compatible with the scheme proposed in Ref.~\cite{barenco}, where
a decomposition of multi-qubit CNOT gates into
a network of two-qubit CNOT gates has been presented.  
However, this decomposition may require many elementary operations. It seems
to be more appropriate for some practical implementations of quantum computing
(for example, computing with 
 cold trapped ions \cite{c&z}) to implement directly multi-qubit CNOT gates.

\section{Quantum logic networks for the state synthesis}
\label{sec3}

In this Section we  present  quantum logic networks
for the synthesis of specific types of coherent superpositions
of multi-qubit quantum  states. 
Later we
will use this result  for construction of an algorithm for a 
generation of an arbitrary pure quantum state of $N$ qubits.

We will consider a quantum register of size $N$, i.e. $N$ qubits. 
Let us denote
\be
\label{*}
|1\r^{N}=\prod\nolimits_{j=1}^N\otimes|1\r_{m_j}\,,\qquad 
|1\r^{N-1}|0\r_{m_k}=
\left(\prod\nolimits_{{j=1 \atop j\neq k}}^N\otimes |1\r_{m_j}\right)
\otimes |0\r_{m_k}\,.
\ee 

Firstly, let us consider a simple network consisting of a multi-qubit
control-$R$ gate having $(N-1)$ control qubits ($c_1,\dots,c_{N-1}$) and
a single target qubit ($t_1$) [see FIG. \ref{obr3}].
Let us assume that all qubits have been initially prepared in the state
$|1\r$, i.e. the whole system is in the state
$|1\r^N$ and the gate realizes the operation
\be
\label{t1}
|1\r^N\longrightarrow R_{01}|1\r^{N-1}|0\r_{t_1}+R_{11}|1\r^N\,,
\ee
where $R_{01}$ and $R_{11}$ are defined by the relation (\ref{2.3}).

\bo[htb]
\centerline{\epsfig{width=2cm, file=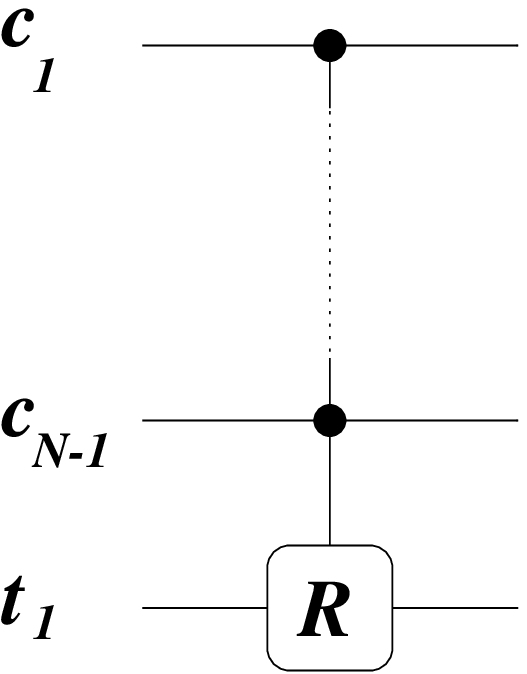}}
\caption{{\footnotesize
The network realizing the transformation given by (\ref{t1}).}}
\label{obr3}
\eo

Secondly, let us consider a network with
$(N-2)$ control qubits
($c_1,\dots,c_{N-2}$) and two target qubits ($t_1,t_2$) [see FIG.~\ref{obr4}].
The network acts on the initial state $|1\r^N$ as follows 
(each arrow in the figure corresponds to an action of a gate in
the sequence)
\be
\label{t2}
|1\r^N&\longrightarrow&
R_{01}|1\r^{N-1}|0\r_{t_1}+R_{11}|1\r^N\nonumber\\
&\longrightarrow&
R_{01}|1\r^{N-2}|0\r_{t_1}|0\r_{t_2}+R_{11}|1\r^{N-1}|0\r_{t_2}  
\nonumber\\
&\longrightarrow&
R_{01}|1\r^{N-2}|0\r_{t_1}|0\r_{t_2}+R_{11}|1\r^N\,.
\ee

\bo[htb]
\centerline{\epsfig{width=3.5cm, file=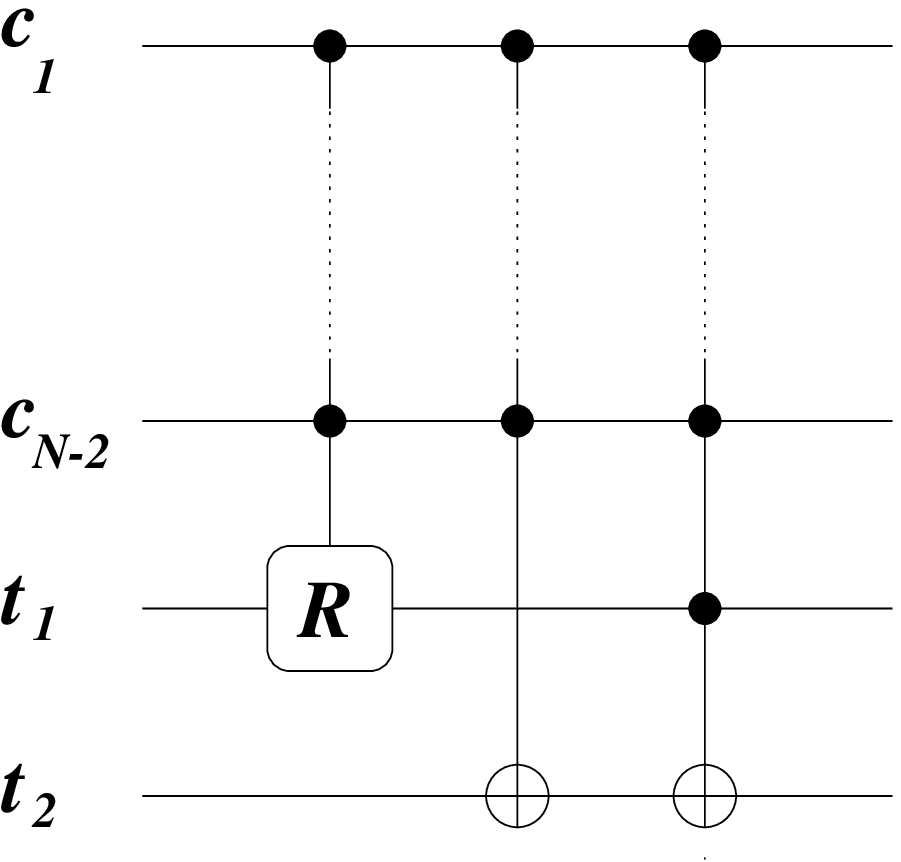}}
\caption{{\footnotesize
The network realizing the transformation  (\ref{t2}).}}
\label{obr4}
\eo

Further, we  design a network with $(N-3)$
control qubits ($c_1,\dots,c_{N-3}$) and three target qubits ($t_1,t_2,t_3$)
[see FIG.~\ref{obr5}]. This network acts as follows 
\be
\label{t3}
|1\r^N&\longrightarrow&
R_{01}|1\r^{N-1}|0\r_{t_1}+R_{11}|1\r^N\nonumber\\
&\longrightarrow&
R_{01}|1\r^{N-2}|0\r_{t_1}|0\r_{t_2}+R_{11}|1\r^{N-1}|0\r_{t_2}\nonumber\\
&\longrightarrow&
R_{01}|1\r^{N-3}|0\r_{t_1}|0\r_{t_2}|0\r_{t_3}+
R_{11}|1\r^{N-2}|0\r_{t_2}|0\r_{t_3}\nonumber\\
&\longrightarrow&
R_{01}|1\r^{N-3}|0\r_{t_1}|0\r_{t_2}|0\r_{t_3}+
R_{11}|1\r^{N-1}|0\r_{t_3}\nonumber\\
&\longrightarrow&
R_{01}|1\r^{N-3}|0\r_{t_1}|0\r_{t_2}|0\r_{t_3}+
R_{11}|1\r^{N}\,.
\ee

\bo[htb]
\centerline{\epsfig{width=5cm, file=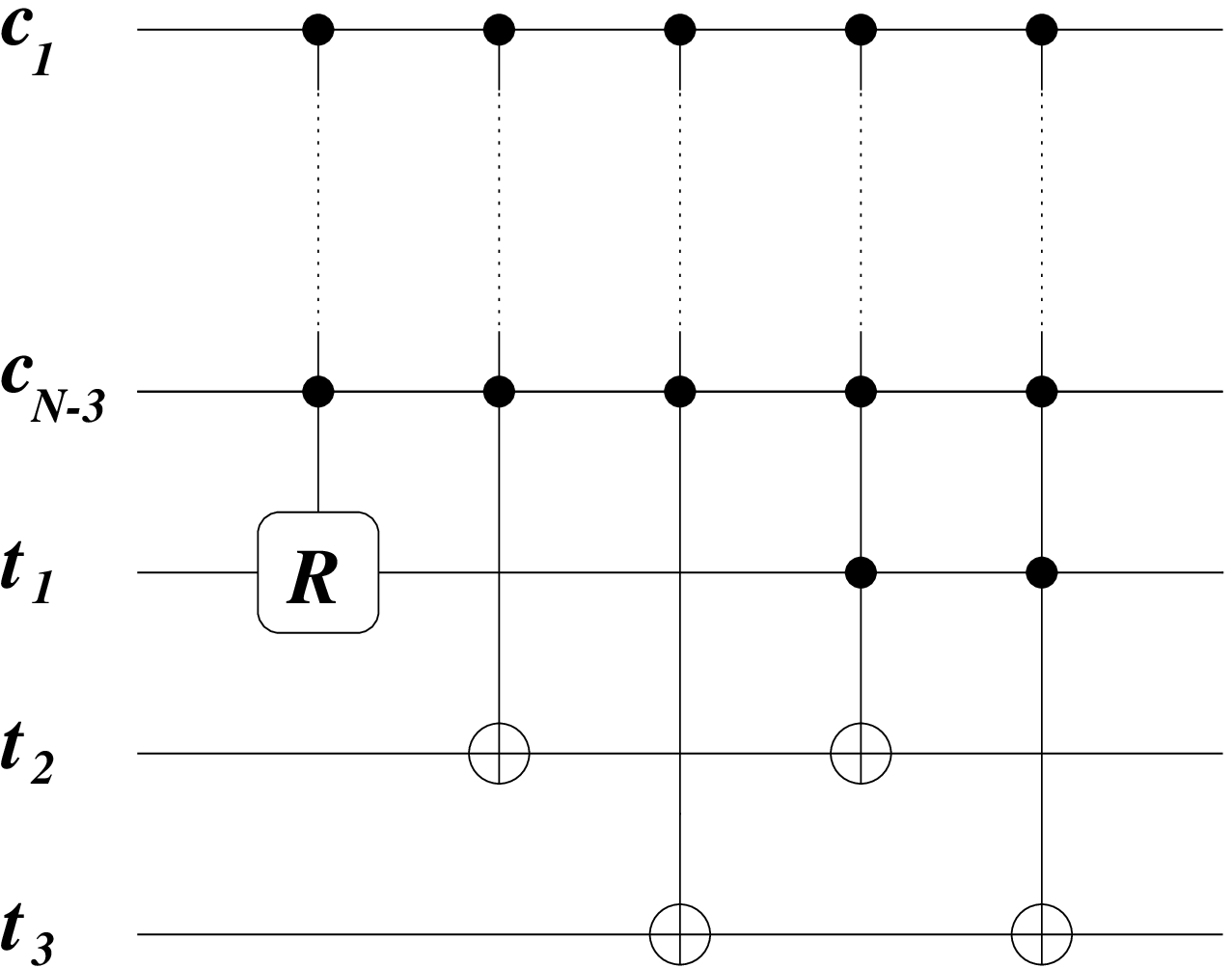}}
\caption{{\footnotesize
The network realizing the transformation (\ref{t3}).}}
\label{obr5}
\eo

The arrangement of quantum logic networks with more target qubits is
straightforward. One has to add another multi-qubit CNOT gate acting 
on the added target
qubit and then one more multi-qubit CNOT gate must be included at 
the end of the network in order 
to erase ``unwanted'' changes on all other terms in a superposition state [for
instance see the $4^{th}$ and $5^{th}$ line in Eq.~(\ref{t3})].

As an example let us consider a network that prepares a pure symmetric (with
respect to permutations) entangled state with just one qubit in the state
$|0\r$ and all others in the state $|1\r$ [see Eq.~(\ref{1.1})].
It can be shown that this state exhibits 
the maximum degree of
entanglement  between any pair of $N$ qubits  \cite{Koashi}. 
The network for the
synthesis of the state (\ref{1.1}) from the initial state $|1\r^N$ is 
shown in 
FIG.~\ref{obr6}, where the rotations $U_j$ are defined as follows
\be
\label{3.2}
U_j=\left(
\begin{array}{cc}
\sqrt{\frac{N-j}{N-j+1}} & \frac{1}{\sqrt{N-j+1}}\\
-\frac{1}{\sqrt{N-j+1}} & \sqrt{\frac{N-j}{N-j+1}}
\end{array}
\right)\,,\quad
j=1,\dots,N-1\,.
\ee

The action of the network in FIG. \ref{obr6} can be described as follows
\be
\label{3.3}
|1\r^N&\stackrel{U_1}{\longrightarrow}&
\frac{1}{\sqrt{N}}|1\r^{N-1}|0\r_1+\sqrt{\frac{N-1}{N}}|1\r^N\nonumber\\
&\stackrel{CU_1}{\longrightarrow}&
\frac{1}{\sqrt{N}}|1\r^{N-1}|0\r_1+\frac{1}{\sqrt{N}}|1\r^{N-1}|0\r_2+
\sqrt{\frac{N-2}{N}}|1\r^N\nonumber\\
&\longrightarrow&
\dots=\frac{1}{\sqrt{N}}\sum_{j=1}^{N-2}|1\r^{N-1}|0\r_j+
\sqrt{\frac{2}{N}}|1\r^N\nonumber\\
&\stackrel{CU_{N-1}}{\longrightarrow}&
\frac{1}{\sqrt{N}}\sum_{j=1}^{N-2}|1\r^{N-1}|0\r_j+
\frac{1}{\sqrt{N}}|1\r^{N-1}|0\r_{N-1}+\frac{1}{\sqrt{N}}|1\r^N\nonumber\\
&\stackrel{CNOT}{\longrightarrow}&
\frac{1}{\sqrt{N}}\sum_{j=1}^{N-1}|1\r^{N-1}|0\r_j+
\frac{1}{\sqrt{N}}|1\r^{N-1}|0\r_N=
\frac{1}{\sqrt{N}}\sum_{j=1}^{N}|1\r^{N-1}|0\r_j\,,
\ee
where $|1\r^N$ denotes the state with all qubits in the state $|1\r$ and
$|1\r^{N-1}|0\r_j$ represents the state of the register with $(N-1)$ qubits
in $|1\r$ and the $j$th qubit in the state $|0\r$ [see the notation
in Eq.~(\ref{*})].

\bo[htb]
\centerline{\epsfig{width=5.5cm, file=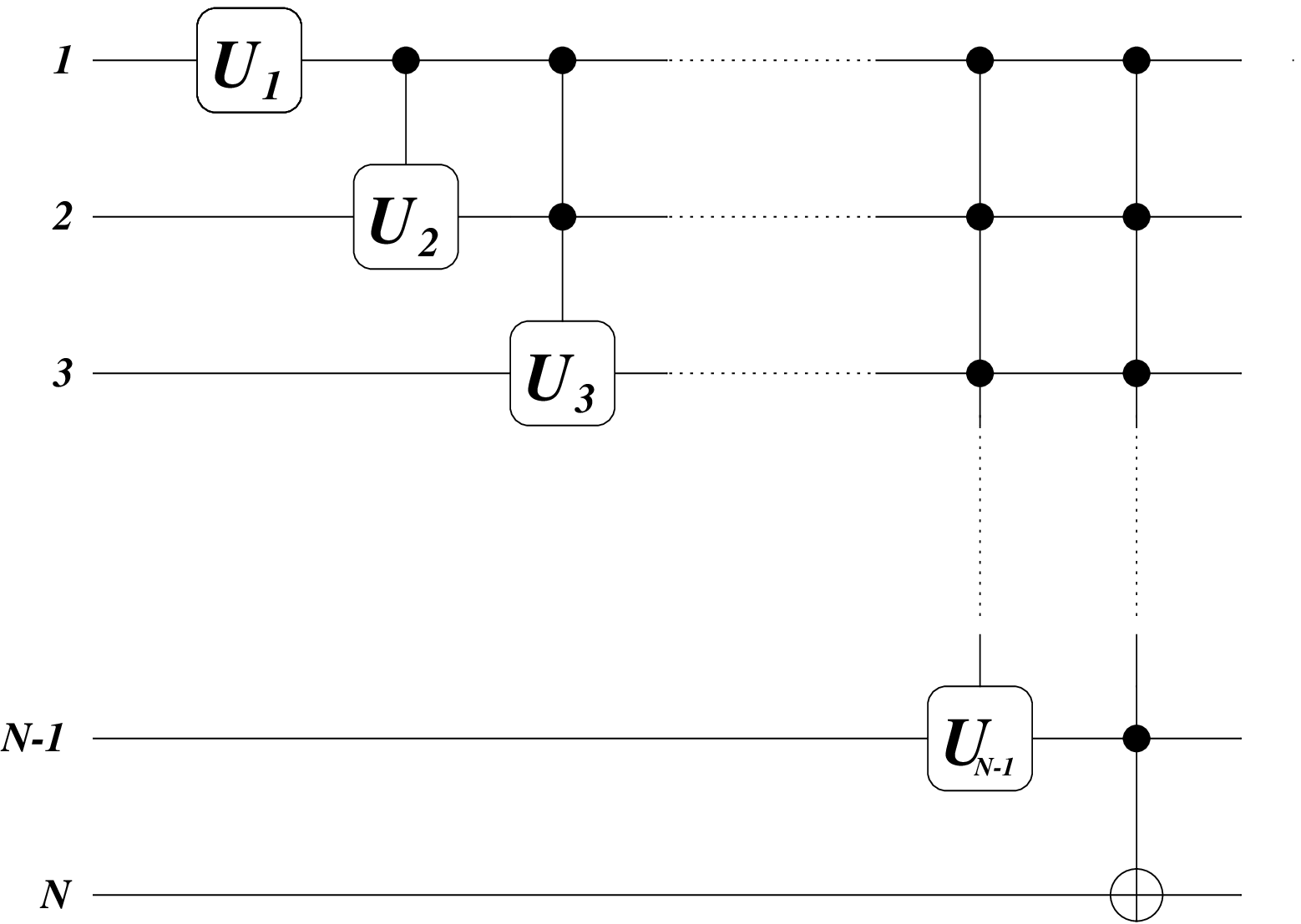}}
\caption{{\footnotesize
The network for the synthesis of the symmetric entangled
state (\ref{1.1}) on $N$ qubits. The rotations $U_j$ are given 
by Eq.~(\ref{3.2}). The $N$ qubits are assumed to be initially in the 
state  $|1\r^N$.}}
\label{obr6}
\eo

A very simple example is the synthesis of the GHZ state, i.e. a
coherent superposition with all qubits to be in the state $|0\r$ or $|1\r$
with the same probability, i.e. $|\Xi\r_{GHZ}=(|0\r^N+|1\r^N)/\sqrt{2}$. 
The corresponding network is shown in FIG. \ref{obr7}. 
The single-qubit rotation $R=O(\pi/2,\pi)$ defined in (\ref{2.3}) is
applied on the initial state $|0\r^N$ and prepares the superposition
$(|0\r^N+|0\r^{N-1}|1\r_1)/\sqrt{2}$. Applying sequentially all CNOT gates one
prepares the GHZ state $|\Xi\r_{GHZ}$.

\bo[htb]
\centerline{\epsfig{width=5cm, file=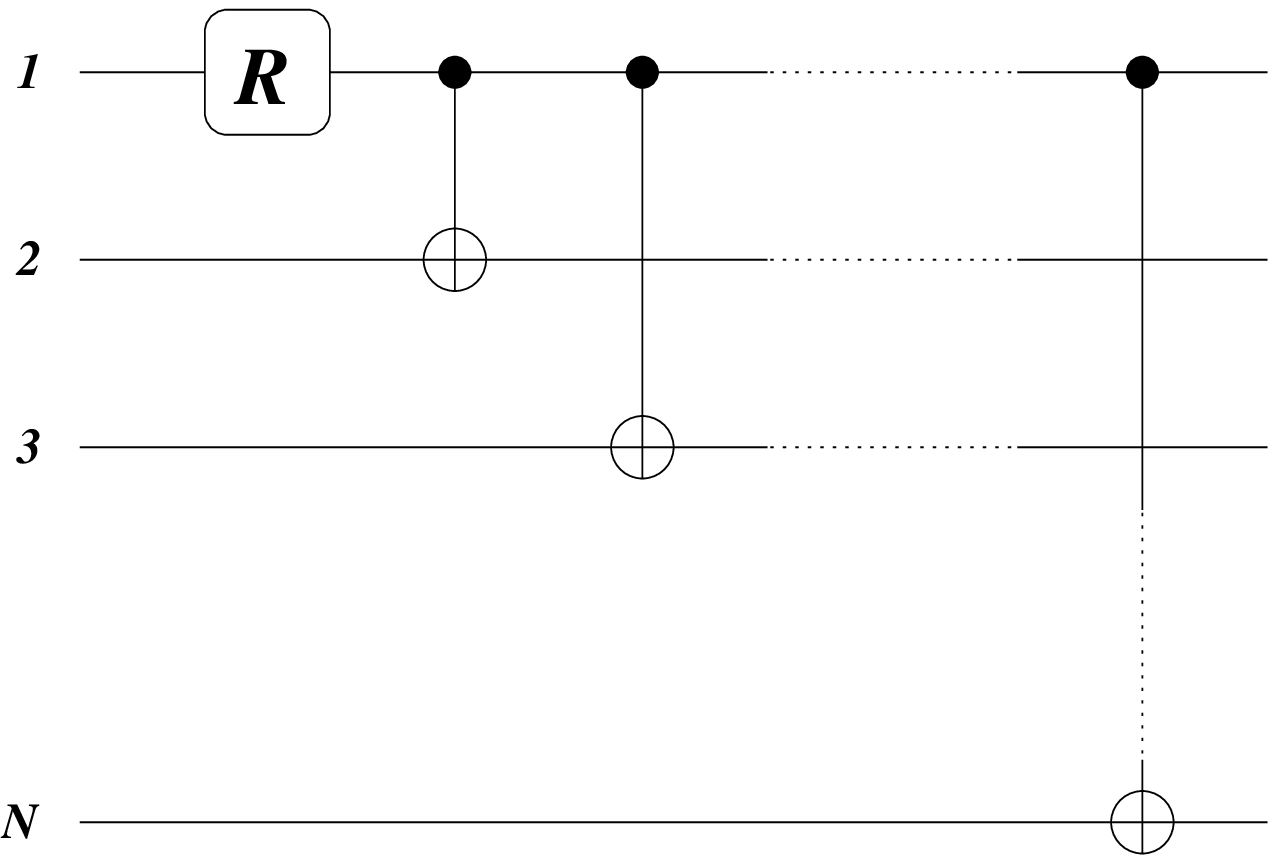}}
\caption{{\footnotesize
The network for the synthesis of the generalization of the 
GHZ state. 
The single-qubit rotation $R$ is given by 
Eq.~(\ref{2.3}) for $R=O(\pi/2,\pi)$. 
The initial state is $|0\r^N$.}}
\label{obr7}
\eo

\section{Synthesis of an arbitrary pure quantum state}
\label{sec4}

Coherent manipulation with states of quantum registers and, in particular,
the synthesis of an arbitrary pure quantum state is of the central
importance for quantum computing. One of the important tasks
is the preparation of  multi-qubit entangled states.

Based on the discussion presented above we can 
propose an array of quantum logic
networks that prepare an arbitrary state from the 
register initially prepared in the state $|0\r^N$, i.e.
\be
\label{4.1}
|0\r^N\quad\longrightarrow\quad
|\psi(N)\r=\sum_{{j=0 \atop x_j\in\{0,1\}^N}}^{2^N-1}c_j|x_j\r=
\sum_{x=00\dots 0}^{11\dots 1}c_x|x\r\, ,
\ee
where $x$ is a binary representation of the number $2^j$.
The proposed scheme can be generalized on the quantum register of an
arbitrary size, but for simplicity 
we will firstly consider the case 
of three qubits.

A general state of three qubits is given as 
\be
\label{4.2}
|\psi(3)\r&=&
\a_0|000\r+e^{i\varphi_1}\a_1|001\r+e^{i\varphi_2}\a_2|010\r+
e^{i\varphi_3}\a_3|100\r\nonumber\\
&+&e^{i\varphi_4}\a_4|011\r+e^{i\varphi_5}\a_5|101\r+
e^{i\varphi_6}\a_6|110\r+e^{i\varphi_7}\a_7|111\r\,,
\ee
where $\a_0,\dots,\a_7$ are real numbers satisfying the normalization
condition
\be
\label{4.3}
\sum_{j=0}^7\a_j^2=1\, ,
\ee
and $\varphi_1,\dots,\varphi_7$ are relative phase factors. 
The global phase is chosen such that $\varphi_0=0$.

\bo[htb]
\centerline{\epsfig{width=9.8cm, file=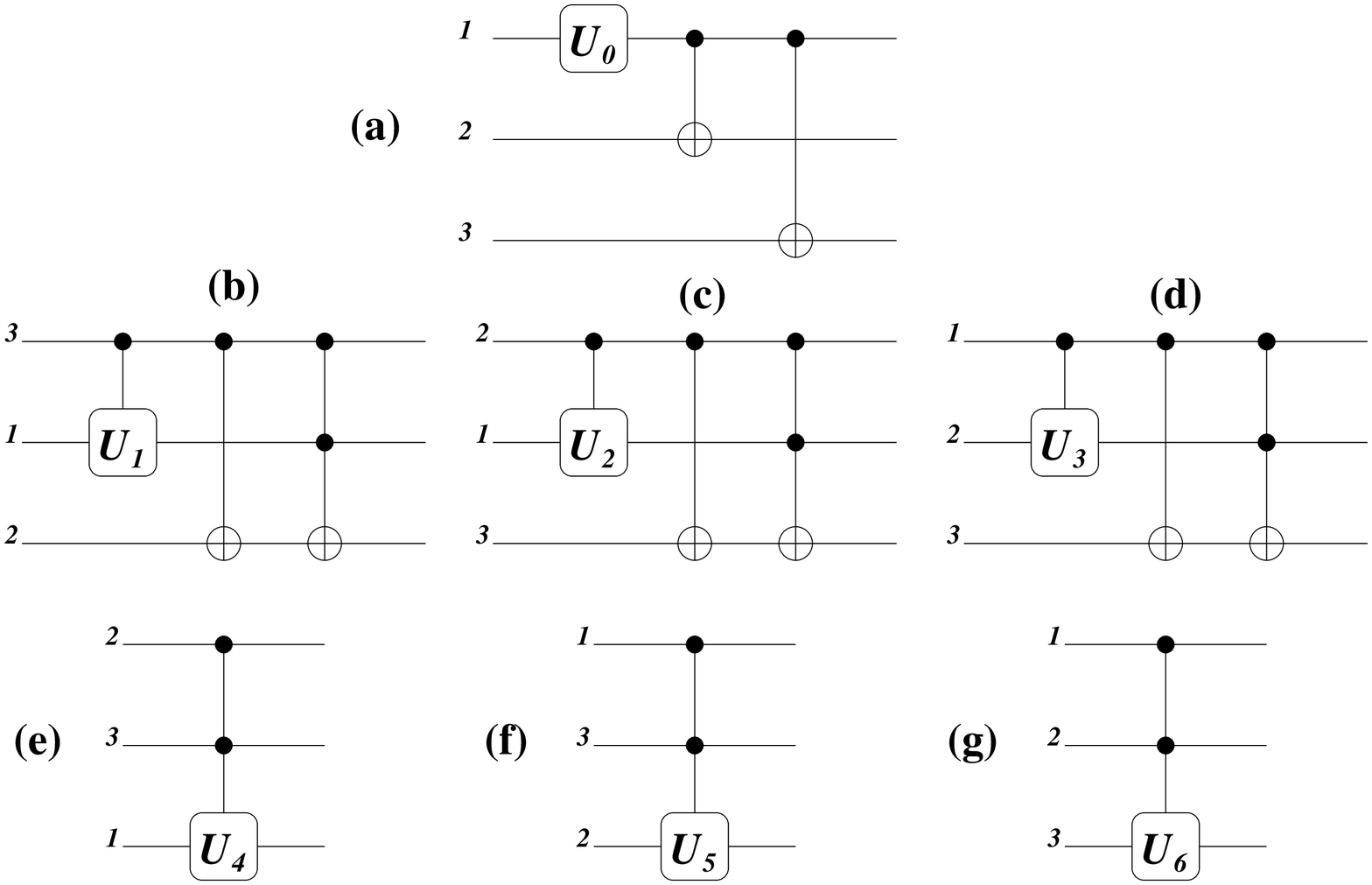}}
\caption{{\footnotesize
An array of  networks for the synthesis of an arbitrary pure
quantum state (\ref{4.2}) on three qubits. The initial state is $|000\r$ and
the rotations $U_j$ are given by Eq.~(\ref{4.4}).}}
\label{obr8}
\eo

\bo[htb]
\centerline{\epsfig{width=10.5cm, file=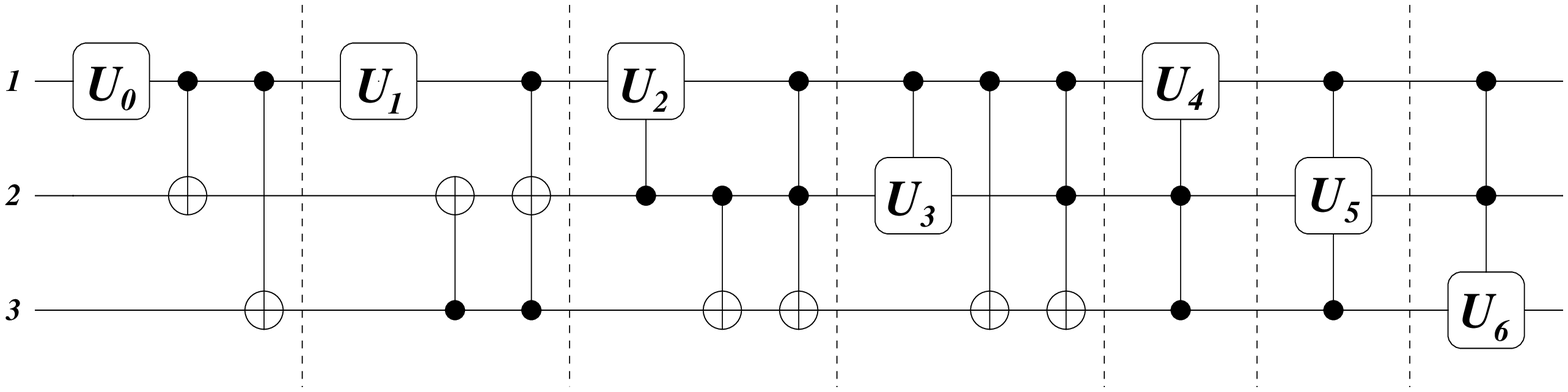}}
\caption{{\footnotesize
A compact form of the array of the networks 
shown in FIG.\,\ref{obr8}.}}
\label{obr8.1}
\eo

In what follows we will present the procedure for the synthesis of the state
(\ref{4.2}). Let us use the abbreviated form of the matrix $R$ defined 
in Eq.~(\ref{2.6}) which we denote as 
\be
\label{4.4}
U_j=\left(
\begin{array}{cc}
a_j & e^{i2\phi_j}b_j\\
-e^{-i2\phi_j}b_j & a_j
\end{array}
\right)\,,\qquad j=0,\dots,6\,,
\ee
where $a_j=\cos\theta_j$ and $b_j=\sin\theta_j$. 
The initial state is $|000\r$. 
The network presented 
in FIG. \ref{obr8} (a) prepares out of the state $|000\r$ 
the superposition
\be
\label{x1}
a_0|000\r-e^{-i2\phi_0}b_0|111\r\,.
\ee
Applying the network in FIG. \ref{obr8} (b), a new term 
\be
\label{x2}
-e^{i2(\phi_1-\phi_0)}b_0b_1|001\r
\ee
is added to the superposition (\ref{x1}) while the
amplitude of the component $|000\r$ is not affected at all. 
The application of the network given by FIG. \ref{obr8} (c)
adds another new term
\be
\label{x3}
-e^{i2(\phi_2-\phi_0)}b_0a_1b_2|010\r
\ee
and does not influence the amplitudes of two foregoing terms $|000\r$ and
$|001\r$. Repeating this procedure, the network in FIG. \ref{obr8} (d) 
adds a new term
\be
\label{x4}
-e^{i2(\phi_3-\phi_0)}b_0b_0a_1a_2b_3|100\r\,.
\ee
Analogously, 
the network shown in  FIG. \ref{obr8} (e) adds a new term
\be
\label{x5}
-e^{i2(\phi_4-\phi_0)}b_0a_1a_2a_3b_4|011\r\,.
\ee
while the networks (f), (g) shown in 
FIG. \ref{obr8} (f) and (g), add  new terms 
\be
\label{x6}
-e^{i2(\phi_5-\phi_0)}b_0a_1a_2a_3a_4b_5|101\r\, , 
\ee
\be
\label{x7}
-e^{i2(\phi_6-\phi_0)}b_0a_1a_2a_3a_4a_5b_6|110\r,
\ee
respectively. The last network shown in  FIG \ref{obr8} (g) also 
determines the amplitude of the last term
\be
\label{x8}
-e^{i2(\phi_7-\phi_0)}b_0a_1a_2a_3a_4a_5a_6|111\r\,.
\ee
Comparing the output from the networks shown 
in FIG. \ref{obr8}, determined by the
relations (\ref{x1})--(\ref{x8}), with the expression (\ref{4.2}), 
we get the final results in TABLE \ref{tab1}.

\begin{table}[htb]
\begin{center}
\begin{tabular}{||c||l|l|c||}
\hline
{\bfseries\itshape \ j\ } & $ \ a_j\ $ & $\ \varphi_j$ & \ state$\ $ \\[0.75mm]
\hline\hline
{\bf 0} & $\ a_0$ & \ 0 (default) & $000$ \\[0.75mm]
\hline
{\bf 1} & $\ b_0b_1$ & $\ 2(\phi_1-\phi_0)+\pi\ $ & $001$ \\[0.75mm]
\hline
{\bf 2} & $\ b_0a_1b_2$ & $\ 2(\phi_2-\phi_0)+\pi\ $ & $010$ \\[0.75mm]
\hline
{\bf 3} & $\ b_0a_1a_2b_3$ & $\ 2(\phi_3-\phi_0)+\pi\ $ & $100$ \\[0.75mm]
\hline
{\bf 4} & $\ b_0a_1a_2a_3b_4$ & $\ 2(\phi_4-\phi_0)+\pi\ $ & $011$ \\[0.75mm]
\hline
{\bf 5} & $\ b_0a_1a_2a_3a_4b_5$ & $\ 2(\phi_5-\phi_0)+\pi\ $ & $101$ \\[0.75mm]
\hline
{\bf 6} & $\ b_0a_1a_2a_3a_4a_5b_6$ & $\ 2(\phi_6-\phi_0)+\pi\ $ & $110$ 
\\[0.75mm]
\hline
{\bf 7} & $\ b_0a_1a_2a_3a_4a_5a_6\ $ & $\ -2\phi_0+\pi$ & $111$ \\[0.75mm]
\hline
\end{tabular}
\end{center}
\caption{{\footnotesize
The network in FIG. \ref{obr8.1} generates the state (\ref{4.2}) from
empty register $|000\r$. The network is characterized by the coefficients $a_j,
\phi_j$, where $b_j=\sqrt{1-a_j^2}$. The state (\ref{4.2}) is determined by
the coefficients $\a_j, \varphi_j$. The table relates these two set of
numbers. The inverse relations are given by the equations (\ref{r1}) and
(\ref{r3}).}}
\label{tab1}
\end{table} 

The coherent superposition (\ref{4.2})
is completely determined by 15 parameters
($\a_0,\dots,\a_7;\varphi_1,\dots,\varphi_7$). The normalization condition
(\ref{4.3}) reduces this number to 14. The networks in FIG. \ref{obr8} 
are determined by 14 parameters ($b_0,\dots,b_6;\phi_0,\dots,\phi_6$). 
Thus, the mapping between the state (\ref{4.2}) and the networks is clearly
defined. From given values of $\a_j$ and $\varphi_j$ one can calculate $b_j$
and $\phi_j$ according to the expressions
\be
\label{r1}
\phi_0=\frac{1}{2}(\pi-\varphi_7)\,, \qquad
\phi_j=\frac{1}{2}(\varphi_j-\varphi_7)\,,\qquad j=1,\dots,6
\ee
and
\be
\label{r3}
b_0=\sqrt{1-\alpha_0^2}\,,\qquad
b_j=\frac{\alpha_j}{\sqrt{1-\sum\limits_{k=0}^{j-1}\alpha_k^2}}\,,\qquad
j=1,\dots,6\,,
\ee
which determine the single-qubit rotations (\ref{4.4}).

The state (\ref{4.2}) contains  terms corresponding to all possible
permutations of three qubits. However, a {\it reduced} superposition with
some terms missing might be desired. For this purpose, we can skip networks
responsible for the synthesis of these terms or 
the corresponding parameter $b_j$ can be
set to zero. 
For instance, in the case when the term $|000\r$ does not appear 
in a final desired quantum state, 
we begin with the initial state
$|111\r$ and skip the network in FIG. \ref{obr8} (a). 
If we do not wish, for a change, 
to generate the term $|111\r$, one may set the parameter $a_6$ to zero and
the phase factor can be chosen arbitrarily (see the table above).

The scheme can be analogically extended to an arbitrary number of qubits.
In what follows 
we will briefly discuss the extension on four qubits. These can be
prepared, in general, in the coherent superposition consisting of 16 terms,
i.e. $|0000\r$, $|0100\r$, $|0010\r$, ... , $|1111\r$.

The network in FIG. \ref{obr9} (a) prepares the superposition 
of the terms $|0000\r$ and
$|1111\r$ with corresponding complex amplitudes depending on the choice of
the single-qubit rotation $R_1$. Application of the network 
in FIG. \ref{obr9} (b) running
through all possible permutations of four qubits, i.e. 
$(c_1,t_1,t_2,t_3)=\{(1,2,3,4);(2,1,3,4);(3,1,2,4);(4,1,2,3)\}$, adds to the
superposition new terms $|1000\r, |0100\r, |0010\r, |0001\r$ with
corresponding amplitudes determined by $R_2$. Further, we apply the network
of the type in FIG. \ref{obr9} (c) running through the permutations
$(c_1,c_2,t_1,t_2)=\{(3,4,1,2);(2,4,1,3);(2,3,1,4);(1,4,2,3);(1,3,2,4);
(1,2,3,4)\}$ and the terms 
$|0011\r,|0101\r,|0110\r,|1001\r,|1010\r,|1100\r$ (with corresponding
amplitudes given by $R_3$) will be included to the state under construction.
Finally, the network in FIG. \ref{obr9} (d) running through 
$(c_1,c_2,c_3,t_1)=\{(2,3,4,1);(3,4,1,2);(4,1,2,3);(1,2,3,4)\}$ generates
new terms $|0111\r,|1011,|1101\r,|1110\r$.

\bo[htb]
\centerline{\epsfig{width=10.3cm, file=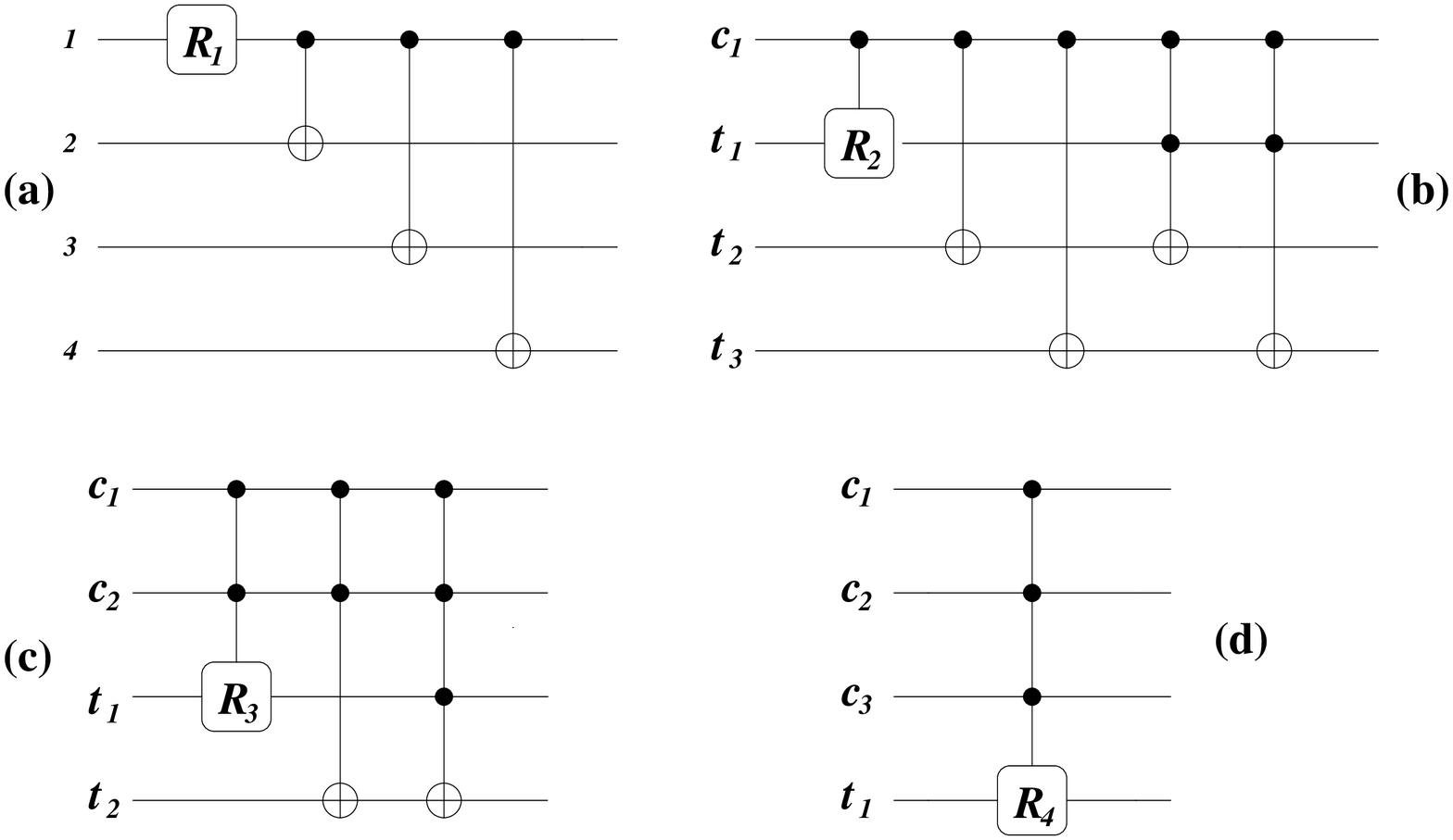}}
\caption{{\footnotesize
An array of  networks for the synthesis of an arbitrary pure
quantum state (\ref{4.1}) of four qubits as discussed in Section
\ref{sec4}).}}
\label{obr9}
\eo     

The extension to $N$ qubits is analogical. 
The state synthesis is started from
the initial state $|0\r^N$. Firstly, one uses the network for the
preparation of  superpositions of $|0\r^N$ and $|1\r^N$ with determined
amplitudes. Secondly, the networks with one control qubit ($c_1$) and $N-1$
target qubits ($t_1,\dots,t_{N-1}$) running through all permutations are
applied. Then, we employ the networks with two control qubits ($c_1,c_2$) and
$N-2$ target qubits ($t_1,\dots,t_{N-2}$). Further, the networks with more
control qubits ($3,4,\dots,N-1$). These procedures are repeated 
until we achieve $N-1$ 
control qubits (and one target qubit). The synthesis stops and a desired final
state is prepared.

\section{Realization on cold trapped ions}
\label{sec5}

In  previous sections we have proposed a scheme for the synthesis of an
arbitrary pure quantum state of a system of $N$ qubits. 
The implementation of the multi-qubit CNOT gate has played the central
role in our scheme. It is well known 
how to decompose multi-qubit gates into
a network of single-qubit and two-qubit CNOT gates \cite{barenco}.
However, it seems that a direct implementation
of multi-qubit CNOT gates in specific quantum
systems is more straightforward and requires less elementary operations
(for example laser pulses) than its decomposition.

We demonstrate this idea on a system of 
{\it cold trapped ions}. We will briefly
describe the system under consideration and  show how multi-qubit gates
can be implemented. 

The quantum system considered here is a model of 
a string of $N$ atomic ions confined
in the linear Paul trap proposed by Cirac and Zoller in 1995
\cite{c&z}. First experiments on a single ion and two ions were realized 
by the NIST group in Boulder \cite{wineland}. Experiments with more ions
were done, for example, by the group in Innsbruck \cite{blatt}.

The confinement of a system of trapped ions 
along the $x$, $y$ and $z$ axis can
be described by an anisotropic harmonic pseudopontential of frequencies
$\omega_z\ll\omega_r$, where for the usual choice of trapping 
radio-frequency (rf) voltage we get
$\omega_r=\omega_x=\omega_y$. The ions are firstly 
Doppler cooled and then undergo
the sideband cooling. Laser cooling minimize their motional energy and the
ions oscillate around their equilibrium positions. In this case we can
describe their motion in terms of normal modes. We will consider only the
lowest, center-of-mass (COM), vibrational collective mode of the
ions along the $z$ axis, when all the ions oscillate back and forth as if
they were a rigid body.
The sideband cooling leaves the ions in the quantum 
ground motional state, therefore we have to assume the Lamb-Dicke limit,
i.e. the photon recoil frequency (corresponding to the laser cooling
transition) is much smaller than the frequency of the considered COM mode. 
The ions in the trap represent qubits with two distinct internal atomic
states denoted as $|g\r$ and $|e\r$ with corresponding energy levels 
$E_g$ and $E_e$, respectively. 
We will consider individual-ion-addressing with
a laser beam of the frequency $\omega_L$ represented by a classical
traveling wave. Then, in the interaction picture, in the rotating wave
approximation plus the weak coupling regime 
and in the Lamb-Dicke limit we can write the Hamiltonian
corresponding to the interaction between the $j$th trapped ion
($j=1,\dots,N$) and the laser
beam tuned on the {\it carrier} ($\omega_L=\omega_0$)
\be
\label{*1}
\hat{{\cal A}}_j=\frac{\hbar\Omega_j}{2}\bigg(|e\r_j\l g|+|g\r_j\l e|\bigg)
\ee
and on the {\it first red sideband} ($\omega_L=\omega_0-\omega_z$)
\be
\label{*2}
\hat{{\cal B}}_j=\frac{\hbar\Omega_j}{2}\frac{i\eta}{\sqrt{N}}
\bigg(|e\r_j\l g|\,\hat{a}+|g\r_j\l e|\,\hat{a}^{\dag}\bigg)\,,
\ee
where $\Omega_j=|\Omega_j|e^{-i\phi}$ is the laser coupling constant, 
$\phi$ is the laser phase, $\eta$ is the Lamb-Dicke parameter, $\hat{a}$
and $\hat{a}^{\dag}$ are the annihilation and creation operator of the
quantized COM mode with the frequency $\omega_z$,
where $\hat{a}^{\dag}\hat{a}|n\r=n|n\r$ and $\omega_0=(E_e-E_g)/\hbar$ is the
atomic transition frequency. 

Further, we can write the unitary evolution
operators via which the action of quantum gates is realized. 
Firstly, let us consider the evolution operator corresponding to
a $k\pi$-pulse on the carrier ($t=k\pi/|\Omega_j|$) 
applied on the $j$th ion with the arbitrary initial choice of the laser 
phase such that
\be
\label{*3}
\hat{A}_j^k(\phi)=
\exp\left[
-\frac{k\pi}{2}\bigg(|e\r_j\l g|\,e^{-i\phi} - |g\r_j\l e|\,e^{i\phi}\bigg)
\right]\, .
\ee
Under the action of this unitary operator the two internal states of
the $j$-th ion are changed as follows
\be
\label{*4}
|g\r_j&\rightarrow&
\cos(k\pi/2)|g\r_j-e^{-i\phi}\sin(k\pi/2)|e\r_j\,,\nonumber\\
|e\r_j&\rightarrow&
\cos(k\pi/2)|e\r_j+e^{i\phi}\sin(k\pi/2)|g\r_j\,.
\ee
Secondly, we have the evolution operator for a $k\pi$-pulse on the first red
sideband ($t=k\pi\sqrt{N}/|\Omega_j|\eta$) on the $j$th ion choosing the
laser phase such that
\be
\label{*5}
\hat{B}_j^{k,q}(\phi)=
\exp\left[
-\frac{ik\pi}{2}\bigg(
|e_q\r_j\l g|\,\hat{a}\,e^{-i\phi}+|g\r_j\l e_q|\,\hat{a}^{\dag}\,e^{i\phi}
\bigg)\right]\,,
\ee
which implies the transformation
\be
\label{*6}
|g\r_j|0\r&\rightarrow&|g\r_j|0\r\,,\nonumber\\
|g\r_j|1\r&\rightarrow&
\cos(k\pi/2)|g\r_j|1\r-ie^{-i\phi}\sin(k\pi/2)|e\r_j|0\r\,,\nonumber\\
|e\r_j|0\r&\rightarrow&
\cos(k\pi/2)|e\r_j|0\r-ie^{i\phi}\sin(k\pi/2)|g\r_j|1\r\,,
\ee
where $q=I,II$ and $|e_I\r$ denotes the upper internal level, whereas
$|e_{II}\r$ refers to an auxiliary internal level $|aux\r$. In the original
proposal \cite{c&z} the values of the parameter 
$q=I,II$ refer to  the situation where the transition excited by
the laser depends on the laser polarization.

The operators (\ref{*3}) and (\ref{*5}) provide us with the
possibility to introduce the implementation of the single-qubit rotation and
multi-qubit CNOT gate on selected ions (representing qubits). 
It is obvious from
the transformation (\ref{*4}) that the evolution operator
(\ref{*3}) corresponds to the single-qubit rotation $O(k\pi,\phi)$ on
the $j$th ion [see the definition (\ref{2.3})]. The two-qubit CNOT gate 
(the $m_1$th ion is the control and the $m_2$th ion is the target) 
is realized by  the evolution operator (from right to left)
\be
\label{*7}
\hat{{\cal A}}_{m_2}^{1/2}(\pi)\,
\hat{{\cal B}}_{m_1}^{1,I}\,
\hat{{\cal B}}_{m_2}^{2,II}\,
\hat{{\cal B}}_{m_1}^{1,I}\,
\hat{{\cal A}}_{m_2}^{1/2}(0)\,, 
\ee
which corresponds to a sequence of pulses as described above. This
transformation acts on two ions as
\be
\label{*8}
|g\r_{m_1}|g\r_{m_2}|0\r & \longrightarrow & |g\r_{m_1}|g\r_{m_2}|0\r\,,
\nonumber\\
|g\r_{m_1}|e\r_{m_2}|0\r & \longrightarrow & |g\r_{m_1}|e\r_{m_2}|0\r\,,
\nonumber\\
|e\r_{m_1}|g\r_{m_2}|0\r & \longrightarrow & |e\r_{m_1}|e\r_{m_2}|0\r\,,
\nonumber\\
|e\r_{m_1}|e\r_{m_2}|0\r & \longrightarrow & |e\r_{m_1}|g\r_{m_2}|0\r\, .
\ee
The ions are assumed to be cooled to the ground vibrational state $|0\r$
before the operation. We have used the notation $\hat{B}\equiv\hat{B}(0)$ in
the relation (\ref{*7}).
The two-qubit CNOT gate can be extended to the multi-qubit
$(\mbox{control})^q$-NOT gate acting on $q+1$ ions ($m_1,\dots,m_q$ ions 
represent the control, while the $m_{q+1}$th ion  is the target) and 
can be realized by the following evolution operator (from right to left)
\be
\label{*9}
\hat{{\cal A}}_{m_{q+1}}^{1/2}(\pi)\,
\hat{{\cal B}}_{m_1}^{1,I}\,
\left[\prod_{j=2}^q\hat{{\cal B}}_{m_j}^{1,II}\right]\,
\hat{{\cal B}}_{m_{q+1}}^{2,II}\,
\left[\prod_{j=q}^2\hat{{\cal B}}_{m_j}^{1,II}\right]\,
\hat{{\cal B}}_{m_1}^{1,I}\,
\hat{{\cal A}}_{m_{q+1}}^{1/2}(0)
\ee
corresponding to the transformation
\be
\label{*10}
|\Psi_{no}\r|g\r_{m_{q+1}}|0\r
&\longrightarrow&
|\Psi_{no}\r|g\r_{m_{q+1}}|0\r\,,\qquad
|\Psi_{no}\r\neq\prod_{j=1}^q\otimes|e\r_{m_j}\,,\nonumber\\
|\Psi_{no}\r|e\r_{m_{q+1}}|0\r
&\longrightarrow&
|\Psi_{no}\r|e\r_{m_{q+1}}|0\r\,,\nonumber\\
|\Psi_{yes}\r|g\r_{m_{q+1}}|0\r
&\longrightarrow&
|\Psi_{yes}\r|e\r_{m_{q+1}}|0\r\,,\qquad
|\Psi_{yes}\r=\prod_{j=1}^q\otimes|e\r_{m_q}\,,\nonumber\\
|\Psi_{yes}\r|e\r_{m_{q+1}}|0\r
&\longrightarrow&
|\Psi_{yes}\r|g\r_{m_{q+1}}|0\r\,.
\ee
It is obvious from Eqs.~(\ref{*8}) and (\ref{*10}) that the ions must be 
kept in the ground motional state. This
arrangement eliminates heating processes which lead to decoherence.
However, it is still the experimental challenge to cool to the ground state
$|n=0\r$ more than two ions.

\section{Discussion and conclusions}
\label{sec6}

In this paper we have shown how  multiparticle entangled states
can be constructed with the help of multi-qubit quantum gates.
We have shown how to implement these gates on the system of cold
trapped ions. This allows us to ``realize''any multi-qubit 
control-$R$ gate and also any logic network proposed 
in Sec. \ref{sec3} and \ref{sec4}.

To understand the feasibility of this algorithm 
we present some  estimations considering the application of
the introduced gates and networks on cold trapped ions. 

The main aim of further discussion is to illustrate a range of
relevant physical parameters for implementation of proposed scheme.
Obviously, specific experimental setups have to be considered
separately. We present just rough estimates of minimal 
times required for realization of desired gate operations.


Let us consider  Calcium ions $^{40}\mbox{Ca}^+$ with 
the ``ground'' (computational) state $|g=S_{1/2}\r$ and 
the ``excited'' (computational) state $|e=D_{5/2}\r$. The lifetime of the ion on
the~metastable $D_{5/2}$ level is 1.045\,s. 

We will assume $N$ ions loaded and confined in the trap. The ions will be
individually addressed with a laser beam $(\lambda=729\mbox{nm})$ supposing
the Gaussian intensity profile $I/I_0=\exp(-2\rho^2/w_0^2)$, where $\rho$
denotes the radial distance and $2w_0=10\mu\mbox{m}$ is the beam waist.
Further, let the angle between the laser beam and the $z$ axis be
$\vartheta=60^{\circ}$. Then, the recoil frequency of the Calcium ion is
$f_R=2.33\mbox{kHz}$, where $f_R=E_R/h$, $E_R=\hbar^2k^2/2m$,
$k=2\pi/\lambda$ and $h=2\pi\hbar$. The axial trapping frequency is
$\omega_z/2\pi=110\mbox{kHz}$. We can also calculate the Lamb-Dicke
parameter $\eta=\sqrt{E_R/\hbar\omega_z}$, i.e. $\eta=0.15$. The minimum
spacing between two neighboring ions is determined by the approximate formula
\cite{steane1, james}:
\be
\label{+}
\Delta z_{min}\simeq\frac{2.018}{N^{0.559}}
\left(\frac{q^2}{4\pi\varepsilon_0m\omega_z^2}\right)^{1/3}\,,
\ee
where $q$ is the ion charge, $m$ is the ion mass and $\epsilon_0$ is the
permitivity of vacuum.

The multi-qubit CNOT gate on the ions is realized by the evolution operator
(\ref{*9}). We will consider three types of elementary operations: (1)
$\pi/2$-pulse on the carrier $({\cal A})$ defined by the relation
(\ref{*3}), (2) $\pi$-pulse $({\cal B}^1)$ and 
(3) $2\pi$-pulse $({\cal B}^2)$ on the first red sideband (\ref{*5}). Each
elementary operation takes a certain time to be implemented on the system of
cold trapped ions. Steane et al. addressed in detail the speed of ion trap
information processors in \cite{steane2}. 

Firstly, the single-qubit rotation $({\cal A})$ can be made much faster than
two-qubit operations $({\cal B}^1, {\cal B}^2)$, because the Lamb-Dicke
parameter $\eta$ can be set zero (i.e., the laser beam 
is perpendicular to the $z$
axis). Thus, $|\Omega|$ can be made large without restrictions on the weak
coupling regime characterized by the condition
$|\Omega|\ll\omega_z$. We will assume
$|\Omega|/2\pi=50\mbox{kHz}$ and estimate the time required for the
single-qubit rotation as $T_{{\cal A}}=\pi/2|\Omega|=5\mu\mbox{s}$.

Secondly, by definition 
for the operations ${\cal B}^1$ and ${\cal B}^2$, the Lamb-Dicke
parameter must be non zero  [see Eq.~(\ref{*2})]. This means that
some  unwanted off-resonant transitions will be present,
which may significantly
affect times required for the operations ${\cal B}^{1,2}$

In Ref.~\cite{steane2} it has been shown 
that the minimal time $T_{{\cal B}}$ for the realization of the
operation ${\cal B}^1$ is proportional to the geometric mean of the recoil
and trapping frequency, i.e.
\be
\label{**}
\frac{1}{T_{{\cal B}}}\simeq\frac{2\sqrt{2}\epsilon}{\sqrt{N}}
\sqrt{\frac{E_R}{h}\frac{\omega_z}{2\pi}}\,,
\ee
where the imprecision $\epsilon=\sqrt{1-F}$ is defined via the fidelity $F$
of the process. The time for the operation ${\cal B}^2$ is then 
$2T_{{\cal B}}$.

Once the gate times are estimated, we can determine the minimal 
total time $T$ required for the experimental 
preparation of the state (\ref{1.1}) on Calcium ions. The total time $T$ is
the sum of times of all operations ${\cal A}, {\cal B}^1, {\cal B}^2$, which
appear in the implementation of the network in FIG. \ref{obr6}. 
The total number of all operations, when preparing the state (\ref{1.1}) 
on $N$ ions, is $2N^2+4N-10$. The explicit expression for the total
time reads
\be
\label{T}
T=N({\cal A})T_{{\cal A}}+N({\cal B}^1)T_{{\cal B}}+
N({\cal B}^2)2T_{{\cal B}}\, .
\ee

In what follows we will consider several situations with the number
of trapped ions varying from 2 to 20. 
In a given ion trap 
for different values of ions we obtain
different minimal spacings $\Delta z_{min}$
 [see Eq.~(\ref{+})]. The minimal spacing between ions
has to be larger than  the half-width of the Gaussian profile of 
the addressing laser beam. In the Innsbruck experiment \cite{xxx}
the width of the Gaussian profile is proportional to $10\mu m$.
Even for 20 ions 
with $\Delta z_{min}=7.31\mu\mbox{m}$ [see Eq.(\ref{+})]
and the given width of the Gaussian profile, 
the ratio between 
the light intensity of the laser addressing a given ion
to  the intensity of the same beam 
  on the neighboring ion is as small as $1.4\%$. 
Therefore,  individual ions can be addressed rather
efficiently.

As follows from Eq.~(\ref{**})
the minimal time for the gate operation depends on the required fidelity
of the process. In our case we consider two values of the fidelity,
namely $F=99\%$ and $F=75\%$. Given these values we can estimate relevant
physical parameters. 

In TABLE \ref{tab2} we present results of our estimations. From here we can
conclude that for a given lifetime of  Calcium ions (1.045 s)
one can perform in our scheme a coherent manipulation with up to
20 ions with the fidelity $99\%$. 
It seems to be a very optimistic
estimation, however we did not optimize the network itself. 

\begin{table}[htb]
\begin{center}
\begin{tabular}{||c||c||c|c||c|c|c||c|c||}
\hline
& &
\multicolumn{2}{|c||}{$T_{{\cal B}}\ [\mu\mbox{s}]$} &
& & & 
\multicolumn{2}{|c||}{$T\ [\mbox{ms}]$}\\[0.75mm] 
\cline{3-4}
\cline{8-9}
{\bfseries\itshape \ N\ } & 
$\ \Delta z_{min}\ [\mu\mbox{m}]\ $ & 
$\ F=99\%\ $ & 
$\ F=75\%\ $ & 
$\ N({\cal A})\ $ & 
$\ N({\cal B}^1)\ $ & 
$\ N({\cal B}^2)\ $ & 
$\ F=99\%\ $ & 
$\ F=75\%\ $\\[0.75mm]
\hline\hline
{\bf 2} & 24.4 & 312 & 62.4 & 3 & 2 & 1 & 1.26 & 0.265\\[0.75mm]\hline
{\bf 3} & 20.8 & 382 & 76.4 & 9 & 8 & 3 & 5.39 & 1.11\\[0.75mm]\hline
{\bf 4} & 18.0 & 441 & 88.3 & 15 & 18 & 5 & 12.4 & 2.55\\[0.75mm]\hline
{\bf 5} & 15.9 & 493 & 98.7 & 21 & 32 & 7 & 22.8 & 4.65\\[0.75mm]\hline
{\bf 6} & 14.3 & 540 & 108 & 27 & 50 & 9 & 36.9 & 7.48\\[0.75mm]\hline
{\bf 7} & 13.1 & 584 & 117 & 33 & 72 & 11 & 55.1 & 11.2\\[0.75mm]\hline
{\bf 8} & 12.2 & 624 & 125 & 39 & 98 & 13 & 77.6 & 15.7\\[0.75mm]\hline
{\bf 9} & 11.4 & 662 & 132 & 45 & 128 & 15 & 105 & 21.1\\[0.75mm]\hline
{\bf 10} & 10.8 & 698 & 140 & 51 & 162 & 17 & 137 & 27.7\\[0.75mm]\hline
{\bf 15} & 8.59 & 855 & 171 & 81 & 392 & 27 & 382 & 76.7\\[0.75mm]\hline
{\bf 20} & 7.31 & 987 & 197 & 111 & 722 & 37 & 786 & 157\\[0.75mm]\hline
\end{tabular}
\end{center}
\caption{{\footnotesize
$N$ is the number of Calcium ions in the trap, $\Delta z_{min}$ is
the minimal distance between two neighboring ions (\ref{+}), $T_{{\cal B}}$
is the minimal time for the realization of the operation 
${\cal B}^1$ [in Eq. (\ref{*9})]
for two different values of the fidelity $(F=99\%, F=75\%)$.
$N({\cal A})$ is the total number of the operations ${\cal A}$ in the
network in FIG. \ref{obr6}, $N({\cal B}^1)$ and $N({\cal B}^2)$ are total
numbers of the operations ${\cal B}^1$ and ${\cal B}^2$, respectively. $T$
is the total minimal time (\ref{T})
for the experimental preparation of the state
(\ref{1.1}) on $N$ ions via the network in FIG. \ref{obr6}.
$T_{{\cal A}}=5\mu\mbox{s}$
is the time for the realization of the operation ${\cal A}$ 
[in Eq. (\ref{*9})].}}
\label{tab2}
\end{table}
 
We have chosen the cold trapped ions as an example for the situation when
the direct implementation of the multi-qubit CNOT gate (using elementary
operations, i.e. in this case laser pulses) is much less demanding than the
decomposition of  multi-qubit CNOT gates into the network of two-qubit CNOT
gates. For instance, let us consider the multi-qubit CNOT gate on six
qubits. Using results of Ref.~\cite{barenco} we can decompose this
multi-qubit CNOT gate into the 
network composed of 12 two-qubit CNOT gates.
In addition, this  network had to be extended by  
three additional auxiliary qubits. 
The multi-bit CNOT gate on $N$ ions (\ref{*9}) is realized by $2N+1$ laser
pulses.
Each two-qubit CNOT gate on two ions 
is then realized using five laser
pulses (\ref{*7}). It means that 
all together 60 pulses have to be used for 12 two-qubit CNOT gates.
However, the direct implementation of the multi-qubit
CNOT gate on six ions 
requires only 13 laser pulses. This difference becomes
even more significant with the increasing number of the ions.
Obviously, smaller the number of pulses easier the scheme can be
implemented.

\acknowledgements
This work was supported by 
the IST projects EQUIP (IST-1999-11053) 
and  QUBITS (IST-1999-13021). We thank Gabriel Drobn\'y for helpful
discussions.    


\end{document}